# *Inverse spin Hall effect in Nd doped SrTiO$_3$*

Qiuru Wang, Wenxu Zhang[a], Bin Peng and Wanli Zhang

State Key Laboratory of Electronic Thin Films and Integrated Devices, University of Electronic Science and Technology of China, Chengdu 610054, People's Republic of China

Conversion of spin to charge current was observed in SrTiO$_3$ doped with Nd (Nd:STO), which exhibited a metallic behavior even with low concentration doping. The obvious variation of DC voltages for Py/Nd:STO, obtained by inverting the spin diffusion direction, demonstrated that the detected signals contained the contribution from the inverse spin Hall effect (ISHE) induced by the spin dependent scattering from Nd impurities with strong spin-orbit interaction. The DC voltages of the ISHE for Nd:STO were measured at different microwave frequency and power, which revealed that spin currents were successfully injected into doped STO layer by spin pumping. The linear relation between the ISHE resistivity and the resistivity induced by impurities implied that the skew scattering was the dominant contribution in this case, and the spin Hall angle was estimated to be $(0.17 \pm 0.05)\%$. This work demonstrated that extrinsic spin dependent scattering in oxides can be used in spintroics besides that in heavy elements doped metals.

## INTRODUCTION

The perovskite-type $3d^0$ oxide SrTiO$_3$ (STO), as an insulator with a wide band gap of about 3.25 eV,[1] has attracted much attention for its potential physical properties, such as superconductivity,[2,3] quantum paraelectricity[4] and ferroelectricity[5]. These excellent physical performances can be introduced by the doping of a small amount of carriers through generating oxygen vacancies[6] or adding dopants such as Cr, La, Nb and Nd[7-10]. In recent decades, due to the increasing interest in spintronics, the spin Hall effect (SHE) in doped materials has been intensively investigated[11-13], which is an important method for converting charge currents into spin currents. In addition to the intrinsic SHE originating from the intrinsic spin-orbit interaction (SOI) in the band structure, the SHE in the doped system is enhanced by the SOI effect in impurity, called the extrinsic SHE. The magnitude of the extrinsic SHE relies on the distinction of the SOI between the host and the impurity. For example, Fert *et al*. found the SHE with large values of spin Hall angle in copper doped with $5d$ heavy metals.[14] Here the intrinsic SHE in Cu is negligibly weak and the SHE signals mainly arise from scattering by impurities presenting strong SOI. Gradhand *et al*. proved the giant SHE in heavy metal Au induced by skew scattering at C and N impurities.[15] The reciprocal effect of SHE is called the inverse spin Hall effect (ISHE), by which spin currents are converted into charge currents as a result of the SOI in nonmagnetic materials as schematically shown in Fig. 1(a). It also provides a way to engineer the magnitude of the ISHE. As reported in the literature, the SOI effect in graphene increased linearly with the impurity coverage.[16] There are two types of mechanisms to take account for this extrinsic effect, namely the skew scattering[17] and the side jump[18]. The

---

a)  Author to whom correspondence should be addressed. E-mail: xwzhang@uestc.edu.cn

former arises from asymmetric scattering from impurities due to the spin-orbit coupling, and the latter can be viewed as a consequence of the anomalous velocity. Normally the ISHE from extrinsic mechanism is observed in metals or semiconductors doped with heavy elements like Ir, Nb, etc. The doped oxides are much less studied, although the controllability of carriers is well demonstrated. The influence to the SOI and SHE is unexplored.

In this work, we present the ISHE measurements on STO substrates lightly doped with nonmagnetic impurities Nd (Nd:STO). The spin current is injected from the ferromagnetic permalloy (Py) film into the adjacent doped STO by spin pumping at the ferromagnetic resonance (FMR), and subsequently transformed into a charge current via the SOI effect induced by the spin skew scattering on the Nd impurities, which can be detected by the shorted microstrip transmission line technique.[19]

### RESULTS AND DISCUSSION

Fig. 1(a) shows a schematic illustration of Py on Nd:STO used in this experiment. The commercially available STO single crystals are cut into size of $0.2 \times 5 \times 10$ mm$^3$, and the doping concentrations of $Nd$ is 0.05 wt%. The ferromagnetic Py films with thickness about 20 nm are deposited on these substrates by magnetron sputtering. During the measurements, the sample is placed at the center of microstrp fixture, where the external magnetic field is applied perpendicular to the direction across the electrodes in the film plane. In the Py/Nd:STO samples, the nonmagnetic layers no longer suppress the further diffusion of the spins due to the introduction of free electrons from impurities, which enables the spin currents to be effectively injected by spin pumping. Thus, the conversion of spin to charge current can be detected via the ISHE. As predicted by Fert, it can be explained by resonant scattering from impurity states split by the SOI, namely the spin Hall angle $\theta_{SH} = 3\lambda_d \sin(2\eta_2 - \eta_1) \sin \eta_1 /(5 \Delta)$ (with $\lambda_d$ being the impurity spin-orbit constant, $\Delta$ being the resonance width, $\eta_1$ and $\eta_2$ being the mean phase shift at the Fermi level).[14] The effect is expected to be larger when the electrons are more localized, where $\Delta$ is smaller, as in the case of doped insulators compared with that in metals. We utilize the two-step measurement with sample flipping to separate the ISHE signals ($V_{ISHE}$) in Nd:STO from the SRE signals ($V_{SRE}$).[20-22] As shown in Fig. 1(b), the DC voltages detected in doped STO are obviously different before ($V_{be}$) and after ($V_{af}$) flipping, which implies that these signals are attributed not only to the contribution from the SRE in Py films, but also to the contribution from the ISHE in doped layers. This significant change arises from the reversal of the sign of $V_{ISHE}$ due to the inverted spin injection by the sample flipping as shown in the inset of Fig. 1(b), and the signals of samples before and after flipping can be respectively described as the addition ($V_{be}$ = $V_{SRE} + V_{ISHE}$) and subtraction ($V_{af} = V_{SRE} - V_{ISHE}$) of the DC voltages of two effects. Thus, the $V_{ISHE}$ is separated from the $V_{SRE}$ through the subtraction of experimental data ($V_{be}$ - $V_{af}$). As depicted in the inset, the line shape of $V_{ISHE}$ is symmetric, which is typical as also shown in previous studies.[23,24]

In contrast, there is no spin current injected into the interface of Py/STO because of the insulating STO suppressing the further diffusion of the spins. Thus, the DC voltage of this sample detected in this case is attributed to the contribution from the SRE in Py film, the line shape of which is a combination of the symmetric and asymmetric Lorentzian components as plotted by the red squares in Fig. 1(c). The SRE, rectifying the microwave current at the FMR, is associated with the precessing magnetization and microwave electric field,[19] but independent on the spin diffusion direction.[25] When the undoped samples are inverted at the steady external magnetic and microwave electric field, the voltage signal of Py/STO is not distinctly different from that of the sample before flipping as depicted by the blue circles in Fig. 1(c).

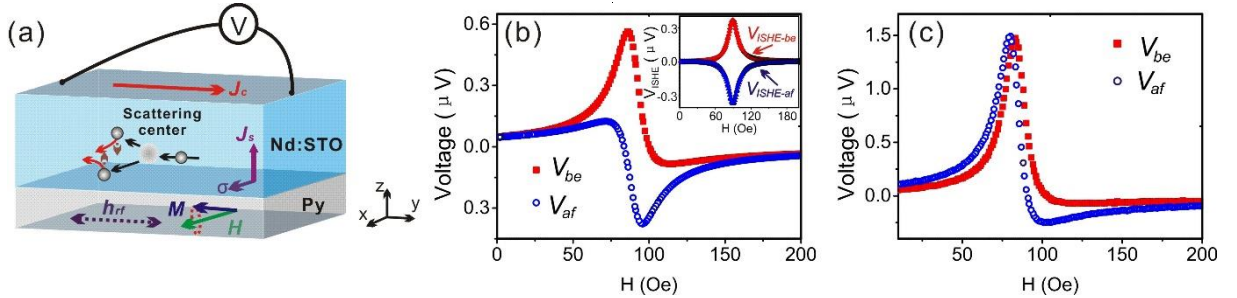

FIG. 1. (a) A schematic illustration of Py/doped STO for the measurement of DC voltages. The magnetic field dependence of the DC voltages of (b) Py/Nd:STO and (c) Py/STO before ($V_{be}$) and after ($V_{af}$) flipping at $f$ = 2.8 GHz, $P$ = 32 mW. The inset in b shows the $V_{ISHE}$ before ($V_{ISHE-be}$) and after ($V_{ISHE-af}$) flipping, respectively.

Fig. 2 shows the external magnetic field dependence of $V_{ISHE}$ for Py/Nd:STO at different microwave power $P$, in which the $V_{ISHE}$ increases linearly with $P$. Furthermore, the values of $V_{ISHE}$ at resonance field $H_r$ for both doped samples are proportional to $P$ as shown in the inset, which is consistent with the model of the DC spin pumping. The spin current density generated by the spin pumping can be expressed as

$$j_s = \frac{G_{\uparrow\downarrow}\gamma^2\hbar h_{rf}^2}{8\pi\alpha^2}\left[\frac{4\pi M_s\gamma + \sqrt{(4\pi M_s\gamma)^2 + 4\omega^2}}{(4\pi M_s\gamma)^2 + 4\omega^2}\right]\frac{2e}{\hbar}, \quad (1)$$

where $G_{\uparrow\downarrow}$, $\gamma$, $\hbar$, $h_{rf}$, $\alpha$, $4\pi M_s$ and $\omega$ denote the spin mixing conductance, gyromagnetic ratio, Dirac constant, amplitude of the microwave magnetic field, Gilbert damping coefficient, effective saturation magnetization and microwave angular frequency, respectively. It indicates that the DC spin current is proportional to the time averaged Gilbert damping term onto the external magnetic field direction, which depends linearly on the square of the magnetization-precession amplitude.[26,27] In terms of the relation between $V_{ISHE}$ and $j_s$, $V_{ISHE} \propto \theta_{SH}j_s \times \sigma$ (with $\theta_{SH}$ being the spin Hall angle, $\sigma$ being the spin-polarization vector),[28] the electric voltage due to the ISHE is also proportional to the square of the amplitude of magnetization-precession, namely the microwave power $P$.[29,30] In our case, the measured linear dependence is consistent with this simplified consideration, which demonstrates the voltages with symmetric Lorentz shape extracted by sample flipping are due entirely to the ISHE induced by the spin pumping.

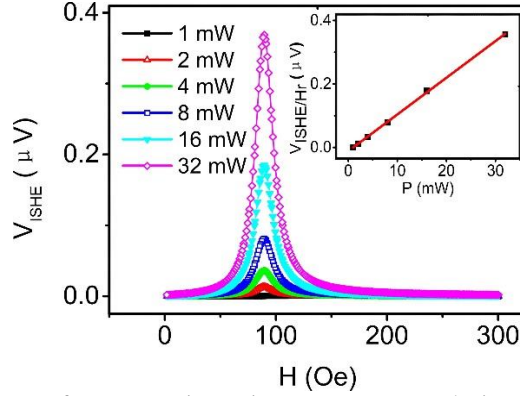

FIG. 2. The magnetic field dependence of $V_{ISHE}$ at various microwave power $P$. The inset shows $P$ dependence of $V_{ISHE}$ at the resonance field $H_r$.

Fig. 3 shows the variation of $V_{ISHE}$ at different microwave frequency ranging from 2.4 to 4.8 GHz, in the step size of 0.4 GHz with fixed $P$ = 32 mW. As depicted in Fig. 3(a), the peak position of voltage curve for doped STO, namely the resonance field $H_r$, increases with the FMR frequency $f$. It is well known that the relation between $H_r$ and $f$ fits the Kittel equation $f = (\gamma/2\pi)\sqrt{H_r(H_r + 4\pi M_s)}$, where $M_s$ is the saturation magnetization of Py. Based on the fitting as plotted by dotted lines in Fig. 3(a), the effective saturation magnetization $4\pi M_s$ for for Py/STO and Py/Nd:STO are determined to be 11.5 kOe and 10.9 kOe, respectively. The difference of this parameter between pristine and doped samples is very small (about 5%), which may arise from the difference in surface roughness of substrates. In addition, it can be clearly seen that voltage curves of the ISHE broaden with increasing the FMR frequency. As shown in Fig.3(b), the FMR linewidth ΔH of doped sample is larger than that of the pristine STO, which experimentally provides the evidence of spin injection induced by the spin pumping[31] The Gilbert damping constant $\alpha$ can be extracted from the function $\Delta H = \Delta H_0 + 4\pi f \alpha/(\sqrt{3}\gamma)$, where $\Delta H_0$ is the inhomogeneous contribution to the linewidth.[32] By fitting the data, we obtain the damping parameters $\alpha_{Py/STO}$ = 0.010 and $\alpha_{Py/Nd:STO}$ = 0.016. The enhanced damping contribution $\Delta \alpha = \alpha_{Py/Nd:STO} - \alpha_{Py/STO}$ is due to the spin flipping. The spin mixing conductance $G_{\uparrow\downarrow}$ can be determined by

$$G_{\uparrow\downarrow} = \frac{4\pi M_s d_F}{g\mu_B} \Delta \alpha, \qquad (2)$$

where $d_F$, $g$, and $\mu_B$ are the thickness of Py film, Landé factor, and Bohr magneton, respectively. Using Eq. (2),

the $G_{\uparrow\downarrow}$ at the interface is estimated as $5.97\times10^{19}$ m$^{-2}$ for Py/Nd:STO, which is in the same order of the values for the Py/Bi ($1.06\times10^{19}$ m$^{-2}$)[33] and Py/n-Ge/Pd samples ($2.15\times10^{19}$ m$^{-2}$)[34], demonstrating that the Py/Nd:STO also enables the efficient spin pumping.

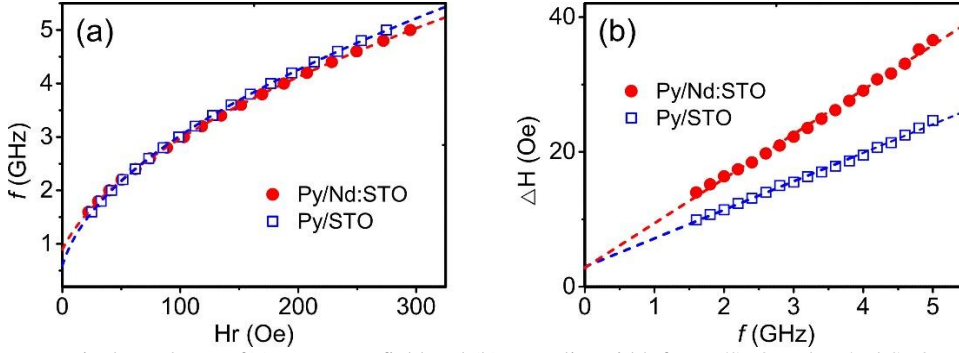

FIG. 3. Ferromagnetic dependence of (a) resonance field and (b) FMR linewidth for Py/STO and Py/Nd:STO, respectively.

Intrinsic STO is an insulator with bandgap larger than 3.0 eV. Fig. 4(a) shows the transport properties of doped STO. The increase in the resistivity with increasing temperature is a typical metallic behavior caused by the introduction of impurities, consistent with the single crystal work of Tufte et al.[35] and Robey et al.[36]. The mobility is 5 cm$^2$V$^{-1}$s$^{-1}$ at room temperature and increases to 90 cm$^2$V$^{-1}$s$^{-1}$ at 80 K. The carrier density extracted from this data is of order $10^{19}$ cm$^{-3}$ and independence on temperature, which can be explained by small donor binding energy of STO due to its large dielectric constant.[37] In this case, an impurity band is formed in doped STO through the overlapping electronic pictures of individual impurity states. When the impurity band overlaps with the conduction band, it results in the transformation from insulator to semiconductor or metal.

The major contributions to the ISHE of doped cases are from the skew scattering or the side jump. In order to figure out the mechanism in Nd:STO, we measured the ISHE resistivity $\rho_{ISHE}$ at different temperatures by $\rho_{ISHE} = w \Delta R_{ISHE} I_C/(xI_S)$, where $w$ is the width of sample, $\Delta R_{ISHE}$ is the amplitude of the ISHE resistance, $I_C$ is the charge current induced by the ISHE, $I_S$ is the effective spin current injected into the Nd:STO and $x$ is a correction factor.[38] As shown in Fig. 4(b), the $\rho_{ISHE}$ is approximately proportional to the resistivity $\rho$ induced by Nd impurities with strong SOI, which provides an indication for the dominant contribution from skew scattering by impurities, compared with the case for the side jump, where the $\rho_{ISHE}$ is proportional to $\rho^2$. The spin Hall angle $\theta_{SH}$, defined as the ratio of $\rho_{ISHE}$ and $\rho$, is estimated as $(0.17\pm0.05)\%$, which is smaller than that for alloys ($\theta_{SH} = 0.6\%$ for Ag doped with Ir, and $\theta_{SH} = 2.1\%$ for CuIr)[39], but larger than that for semiconductors, such as $\theta_{SH} = 0.01\%$ for p-Si[30] and $\theta_{SH} = 0.02\%$ for n-GaAs[40]. The enhancement may come from the extrinsic spin dependent scattering, where the intrinsic contribution is tiny in STO.

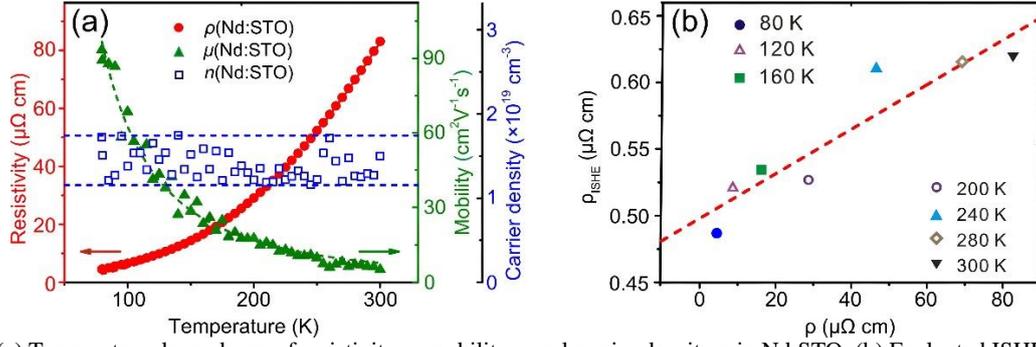

FIG. 4. (a) Temperature dependence of resistivity $\rho$, mobility $\mu$ and carrier density $n$ in Nd:STO. (b) Evaluated ISHE resistivity $\rho_{ISHE}$ as a function of the resistivity $\rho$ induced by impurities.

## CONCLUSIONS

In conclusion, we have shown that doping with Nd leads the STO to the transformation from insulator to metal. By the method of the two-step measurement with sample flipping, the ISHE, giving rise a conversion of spin to charge current caused by the spin skew scattering from impurities, is observed in Nd:STO and separated from the SRE in terms of their different relationship with the direction of spin diffusion. According to the $V_{ISHE}$ signals detected at different microwave frequency and power, we obtain the spin Hall angle of Nd:STO being (0.17±0.05)%, which reflects that spin currents are efficiently injected from Py films into the doped STO, driven by spin pumping. The work provides another paradigm in the oxide spintronics.

## ACKNOWLEDGMENTS

This work was funded by the National High Technology Research and Development Program ("863"-projects, No. 2015AA03130102) and the National Natural Science Foundation of China (NSFC, No. 61471095).

## REFERENCES

[1] K. van Benthem, C. Elsasser, and R. H. French, J. Appl. Phys. 90, 6156 (2001).

[2] J. Schooley, W. Hosler, and M. L. Cohen, Phys. Rev. Lett. 12, 474 (1964).

[3] H. Suzuki, H. Bando, Y. Ootuka, I. H. Inoue, T. Yamamoto, K. Takahashi, and Y. Nishihara, J. Phys. Soc. Jpn. **65**, 1592 (1996).

[4] T Hasegawa, M Shirai, and K Tanaka, J. Lumin. 87, 1217 (2000).

[5] J. H. Haeni, P. Irvin, W. Chang, R. Uecker, P. Reiche, Y. L. Li, S. Choudhury, W. Tian, M. E. Hawley, B. Craigo, A. K. Tagantsev, X. Q. Pan, S. K. Streiffer, L. Q. Chen, S. W. Kirchoefer, J. Levy, and D. G. Schlom, Nature 430, 758 (2004).

[6] D. A. Freedman, D. Roundy, and T. A. Arias, Phys. Rev. B 80, 064108 (2009).

[7] M. Janousch, G. I. Meijer, U. Staub, B. Delley, S. F. Karg, and B. P. Andreasson, Adv. Mater. 19, 2232 (2007).

[8] D. Kan, R. Kanda, Y. Kanemitsu, Y. Shimakawa, M. Takano, T. Terashima, and A. Ishizumi, Appl. Phys. Lett.

88, 191916 (2006).

[9] S. Ohta, H. Ohta, and K. Koumoto, J. Ceram. Soc. Jpn,114, 1325 (2006).

[10] S. W. Robey, V. E. Henrich, C. Eylem, and B. W. Eichhorn, Phys. Rev. B 52, 2395 (1995).

[11] Y. K. Kato, R. C. Myers, A. C. Gossard, and D. D. Awschalom, Science 306, 1910 (2004).

[12] W. K. Tse, and S. D. Sarma, Phys. Rev. Lett. 96, 056601 (2006).

[13] F. Bottegoni, A. Ferrari, G. Isella, M. Finazzi, and F. Ciccacci, Phys. Rev. B 88, 121201(R) (2013).

[14] A. Fert, and P. M. Levy, Phys. Rev. Lett. 106, 157208 (2011).

[15] M. Gradhand, D. V. Fedorov, P. Zahn, and I. Mertig, Phys. Rev. Lett.104, 186403 (2010).

[16] A. H. C. Neto, and F. Guinea, Phys. Rev. Lett. 103, 026804 (2009).

[17] H.-A. Engel, B. I. Halperin, and E. I. Rashba, Phys. Rev. Lett. 95, 166605 (2005).

[18] L. Vila, T. Kimura, and Y. Otani, Phys. Rev. Lett. 99, 226604 (2007).

[19] W. T. Soh, B. Peng, G. Chai, and C. K. Ong, Rev. Sci. Instrum. 85, 026109 (2014).

[20] W. Zhang, B. Peng, F. Han, Q, Wang, W. T. Soh, C. K. Ong, and Wanli Zhang, Appl. Phys. Lett. 108, 102405 (2016).

[21] W. Zhang, Q. Wang, B. Peng, H. Zeng, W. T. Soh, C. K. Ong, and Wanli Zhang, Appl. Phys. Lett. 109, 262402 (2016).

[22] Q. Wang, Wanli Zhang, B. Peng, and W. Zhang, Sol. Stat. Commun. 245 (2016).

[23] L. Bai, Z. Feng, P. Hyde, H. F. Ding, and C. M. Hu, Appl. Phys. Lett. 102, 242402 (2013).

[24] M. Harder, Z. X. Cao, Y. S. Gui, X. L. Fan, and C.-M. Hu, Phys. Rev. B 84, 054423 (2011).

[25] L. Bai, P. Hyde, Y. S. Gui, C. M. Hu, V. Vlaminck, J. E. Pearson, S. D. Bader, and A. Hoffmann, Phys. Rev. Lett. 111, 217602 (2013).

[26] H. Y. Inoue, K. Harii, K. Ando, K. Sasage, and E. Saitoh, J. Appl. Phys. 102, 083915 (2007).

[27] K. Ando, S. Takahashi, J. Ieda, Y. Kajiwara, H. Nakayama, T. Yoshino, K. Harii, Y. Fujikawa, M. Matsuo, S. Maekawa, and E. Saitoh, J. Appl. Phys. 109, 103913 (2011).

[28] H. L. Wang, C. H. Du, Y. Pu, R. Adur, P. C. Hammel, and F. Y. Yang, Phys. Rev. B 88, 100406(R) (2013).

[29] R. Takahashi, R. Iguchi, K. Ando, H. Nakayama, T. Yoshino, and E. Saitoh, J. Appl. Phys. 111, 07C307 (2012).

[30] K. Ando, and E. Saitoh, Nat. Commun. 3, 629 (2012).

[31] S. Kim, D. J. Kim, M. S. Seo, B. G. Park, and S. Y. Park, J. Appl. Phys. 117, 17D901 (2015).

[32] S. S. Kalarickal, P. Krivosik, M. Z. Wu, C. E. Patton, M. L.Schneider, P. Kabos, T. J. Silva, and J. P. Nibarger, J. Appl. Phys. 99, 093909 (2006).

[33] D. Hou, Z. Qiu, K. Harii, Y. Kajiwara, K. Uchida, Y. Fujikawa, H. Nakayama, T. Yoshino, T. An, K. Ando, X. Jin, and E. Saitoh, Appl. Phys. Lett. 101, 042403 (2012).

[34] S. Dushenko, M. Koike, Y. Ando, T. Shinjo, M. Myronov, and M. Shiraishi, Phys. Rev. Lett. 114, 196602 (2015).

[35] O. N. Tufte, and P. W. Chapman, Phys. Rev. 155, 796 (1967).

[36] S. W. Robey, V. E. Henrich, C. Eylem and B.W. Eichhorn, Phys. Rev. B 52, 2395 (1995).

[37] A. Spinelli, M. A. Torija, C. Liu, C. Jan, and C. Leighton, Phys. Rev. B 81, 155110 (2010).

[38] S. Takahashi and S. Maekawa, Sci. Tech. Adv. Mater. 9, 014105 (2008).

[39] Y. Niimi, M. Morota, D. H. Wei, C. Deranlot, M. Basletic, A. Hamzic, A. Fert, and Y. Otani, Phys. Rev. Lett. 106, 126601 (2011).

[40] Y. K. Kato, R. C. Myers, A. C. Gossard, and D. D. Awschalom, Science 306, 1910 (2004).